\documentclass[nofootinbib,twocolumn,superscriptaddress,showpacs,notitlepage]{revtex4-1}
\usepackage{amsmath,amssymb}
\usepackage{graphicx}
\usepackage{txfonts}
\usepackage{color}
\usepackage{hyperref}
\usepackage{bbold}
\usepackage{framed}
\usepackage{wrapfig}
\usepackage[a4paper]{geometry}
\global\long\def\av#1{\left\langle #1 \right\rangle }

\global\long\def\hn{\hat{n}}

\global\long\def\hc{\hat{c}}
\global\long\def\hcd{\hat{c}^\dag}

\begin{document}

\title{
Droplet formation in a one-dimensional system of attractive spinless fermions.
}
\author{Vera V. Vyborova}
\email{vv.vyborova@physics.msu.ru}
\affiliation{Russian Quantum Center, Novaya 100, 143025 Skolkovo, Moscow Region, Russia}
\affiliation{Department of Physics, Lomonosov Moscow State University, Leninskie gory 1, 119991 Moscow, Russia}
\author{Oleg Lychkovskiy}
\affiliation{Skolkovo Institute of Science and Technology,
Nobel street 3, Moscow  121205, Russia}
\affiliation{Steklov Mathematical Institute of Russian Academy of Sciences,
Gubkina str., 8, Moscow 119991, Russia}
\author{Alexey N. Rubtsov}
\affiliation{Russian Quantum Center, Novaya 100, 143025 Skolkovo, Moscow Region, Russia}
\affiliation{Department of Physics, Lomonosov Moscow State University, Leninskie gory 1, 119991 Moscow, Russia}

\begin{abstract}
A translation invariant one-dimensional system of spinless fermions with a finite-range attraction experiences a quantum phase transition to a phase-separated state. While being a conventional Luttinger liquid for a small interaction strength, spinless fermions  form a droplet with the size smaller than the available one-dimensional volume when the interaction strength exceeds some critical value. A particularly remarkable signature of the droplet formation is the change in the lower edge of the many-body excitation spectrum. In the homogeneous phase, it has a Luttinger-liquid shape of periodic arcs on top of the shallow parabolic dispersion of the center-of-mass. When the interaction strength is increased, the arcs disappear completely as soon as the droplet is formed. We perform an exact diagonalization study of this system with the focus on the signatures of the quantum phase transition and the droplet properties. The one-particle and density-density correlation functions, the pressure, the sound velocity, and the droplet density are examined. The value of the critical interaction strength obtained from numerical data reasonably agrees with a simple mean-field analytical estimate. Due to the boson-fermion correspondence valid in one dimension, our results also hold for hard-core bosons with a finite-range attraction.

\end{abstract}

\maketitle

\section{Introduction}
For a long time, superfluid helium has provided the only type of experimentally accessible droplets of a quantum liquid. They proved to be a rich object of experimental and theoretical  studies with interesting physics and important applications~\cite{volovik2003universe}, including molecular spectroscopy and chemistry of matter embedded in the droplets \cite{toennies2004superfluid}.
Recently droplets of quantum liquid were observed in a number of cold atom experiments \cite{schmitt2016self,ferrier-barbut2016observation,cheiney2018bright,cabrera2018quantum}, largely motivated by theoretical proposals of Refs. \cite{bulgac2002dilute,petrov2015quantum}. These experiments triggered an upsurge of interest in phase separation, phase boundaries and droplet formation in the context of quantum fluids \cite{wachtler2016ground-state,macia2016droplets,sekino2018quantum,astrakharchik2018dynamics}.
While the focus of theoretical studies is now on the three-dimensional case already realized experimentally \cite{wachtler2016ground-state,macia2016droplets}, quantum droplets in one dimension (1D) have also been analyzed in recent \cite{sekino2018quantum,astrakharchik2018dynamics} and earlier  \cite{nakamura1997phase,cabra2003instabilities,kolomeisky2003ground-state,kolomeisky2005quantum,law2008quantum,zolner2011ground,yan2014long,edler2015quantum} works.


One-dimensional quantum fluids are conventionally described by the Luttinger liquid theory \cite{haldane1981luttinger}, irrespective of the sign of interaction potential.
It should be stressed that the notion of liquid here emphasizes a substantiation effect of interaction without forming a long-range order. However, it has nothing to do with the thermodynamic terminology on liquids and gases, which implies that in a free space (or a sufficiently large
container)  the liquid  forms a self-bound droplet with a finite density and boundaries, while
the gas expands indefinitely. One-dimensional Fermi systems with repulsion behave as a gas in this sense.
In contrast,  spinless fermions for sufficiently large attraction can form a droplet, and therefore can behave as a liquid in the thermodynamical sense. In the present paper, we investigate this phenomenon by exploring a quantum phase transition to the phase-separated state in a translation invariant 1D system of spinless fermions with a finite-range attraction. \par
While considering phase transitions in low dimensions, one should keep in mind a number of no-go statements and clearly identify domains of their applicability. One statement is that fluctuations destroy a long-range order in low dimension so that an ordered phase cannot exist in a 1D system with linear quasiparticle spectrum even at zero temperature \cite{pitaevskii1991uncertainty}. This prevents phase transition scenarios related to the formation of such a phase.
However, it does not mean that {\it any} phase transition in 1D is forbidden. The role of fluctuations can be downplayed by an external lattice and/or random potential.
For example, a finite temperature phase transition related to the formation of many-body localized phase has been identified in a disordered  1D chain \cite{aleiner2009finite}. Another important statement is the Landau's entropical argument \cite{landau1980statistical}, which suggests that classical statistical physics leaves no space for a phase separation in 1D systems with finite-range interactions. This means that the droplet formation for systems in thermodynamical limit can be realized only at zero temperature.

\par

A possibility of the phase separation in 1D systems was discussed in a number of previous works. It has been known for quite a long time that the Luttinger parameter can diverge  at a sufficiently strong attractive interparticle interaction signifying the quantum phase transition to a phase apparently beyond the Luttinger liquid paradigm \cite{nakamura1997phase,cabra2003instabilities,kolomeisky2003ground-state,kolomeisky2005quantum}. However while for 1D lattice systems the phase separation at zero temperature is a well-studied phenomenon \cite{Giamarchi2003,sutherland2004beautiful,cazalilla2011one, yin2017majorana}, its counterpart
in translation invariant 1D systems is less explored \cite{law2008quantum}. For the latter case,  such a transition has been clearly identified as a formation of a droplet for the first time in Ref. \cite{law2008quantum},
where the variational ansatz has been used to demonstrate the stability of a phase-separated phase. Another important prediction of
\cite{law2008quantum} is that the droplet formation occurs via a second-order transition.
To our knowledge, these findings did not receive much further development, in particular, exact or numerically exact treatment has been lacking up to date.  The present paper aims at filling this gap.

\par




\par  
In this manuscript,  we present a detailed numerical study of the quantum phase transition leading to the formation of the droplet in the translational invariant 1D system of spinless fermions. By employing exact diagonalization we are able to explore the strong coupling regime where the phase transition takes place.
We find that the phase separation is clearly signaled by a qualitative change in the dependence of the system's energy on its total momentum.
We compute a range of physical quantities: two-point and density-density correlation functions, the pressure, the sound velocity and the density of the droplet. Examination of these quantities as a function of interaction strength or/and one-dimensional volume reveal several complimentary signatures of the phase transition.
The Appendix presents a world-line interpretation of the observed phenomenon.

\section{Model and results}
\subsection{Model}
We consider a one-dimensional system of $N$ spinless fermions with a finite-range attractive potential, described by the Hamiltonian
\begin{equation}\label{H}
\hat{H}=- \!\int\! \hcd_x \frac{\partial_x^2}{2 m}  \hc_x\, dx + \frac{1}{2}\!\iint \!V(x-x') \,\hn_x \hn_{x'}\, dx dx',
\end{equation}
where $x,x'$ are defined on the interval  $\left[-\frac{L}{2},\frac{L}{2}\right]$ with periodic boundary conditions imposed.

We have performed an exact diagonalization (ED) study of the system with an attractive potential
\begin{equation}\label{Ur}
	V(x)=\begin{cases} -U \exp\left(-\frac{1}{1-x^2/b^2 }\right), & |x|\le b,~~~~~~~~~~U>0,\\0, & b<|x|\leq\frac{L}{2}\end{cases}
\end{equation}
which is finite-ranged simultaneously in $x$ and $k$ domains. In calculations discussed in this section we have used $b= 1.5 a$, where  $a\equiv L/N$ is the average distance between the particles. Since $V(x)$ is continuous with all its derivatives, its Fourier harmonics $V_k=\int_{-L/2}^{L/2} V(x)\, \cos(k x)\, dx$ drop exponentially fast as $k$ increases. This facilitates numerical calculations in the momentum space, where the Hamiltonian reads
\begin{align} \label{H3}
 \!H\!=\!\sum_{k} \frac{k^2}{2m}\hcd_k\hc_k+\!\frac{1}{2L}\!\!\sum_{q>0,p,k}\!\!V_q\,\hcd_{k+q}\hc_k\hcd_{p-q}\hc_p\!+\!\frac{N^2V_{q=0}}{4L},
\end{align}
momenta $q,\,p,\, k$ being discretized with a step $\delta k=\frac{2 \pi}{L}$ determined by the boundary conditions. Here and after throughout the paper $\hbar=1$. The Fock basis size was limited by a finite number of the one-particle $k$-states. We also assume that the system contains an odd number $N$ of particles, so in the absence of interactions ($U=0$) its ground state at zero total momentum  corresponds to filling all single-particle $k$-states  up to $|k|=k_F-\frac12\delta k$, where the Fermi momentum is $k_F\equiv\pi/a= \frac12N\delta k$.  We measure $U$ in units of the Fermi energy $E_F\equiv k_F^2/2m$ and do not explicitly specify this unit in the legends of plots in order not to overload them. 

\subsection{Spectral edge and elementary excitations \label{subsec spectral edge}}

\begin{figure}[t]
	\begin{minipage}[h]{0.94\linewidth}
		\center{\includegraphics[width=\linewidth]{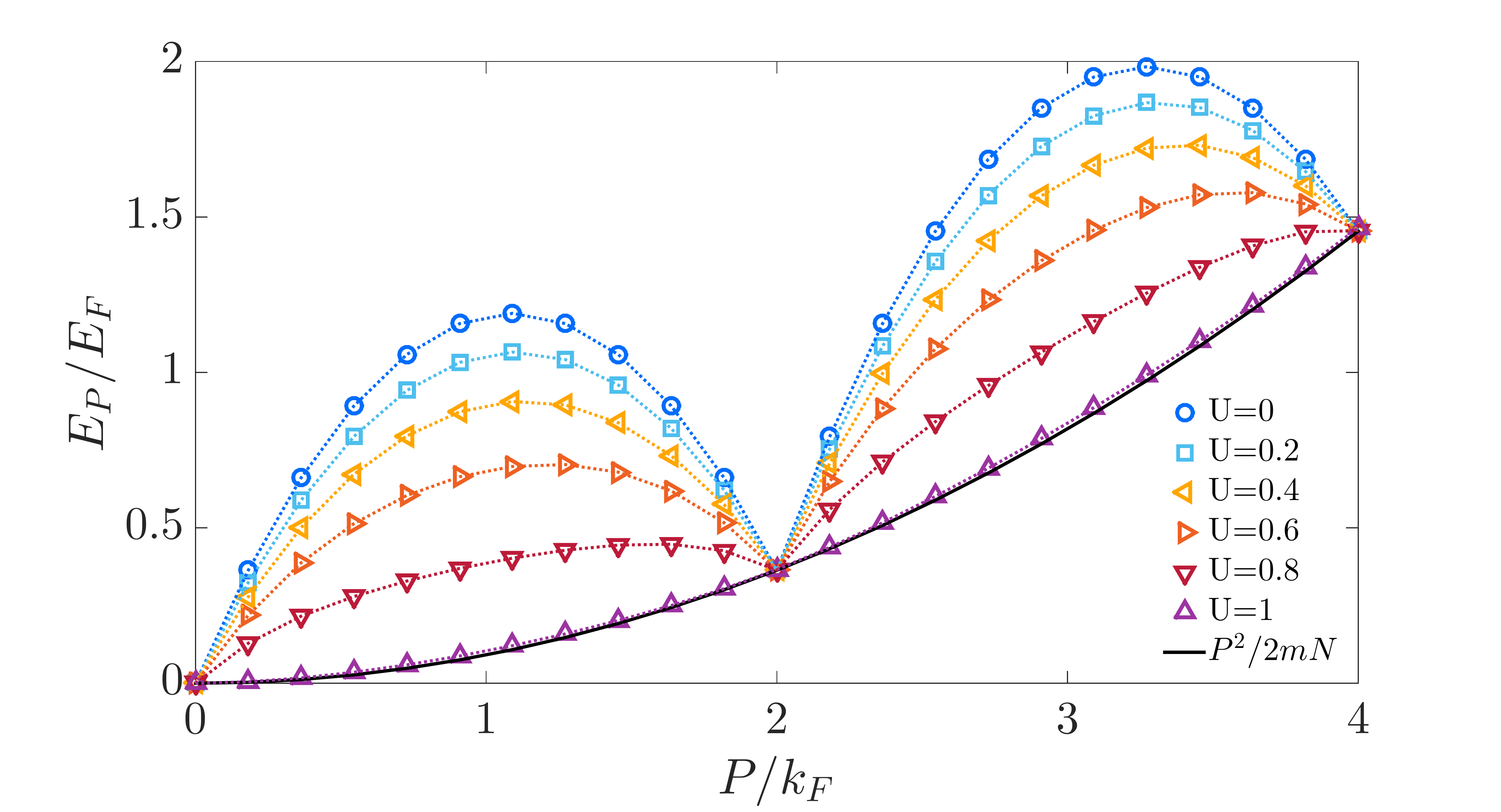}} a) \\
	\end{minipage}
	\vfill
	\begin{minipage}[h]{0.94\linewidth}
		\center{\includegraphics[width=1\linewidth]{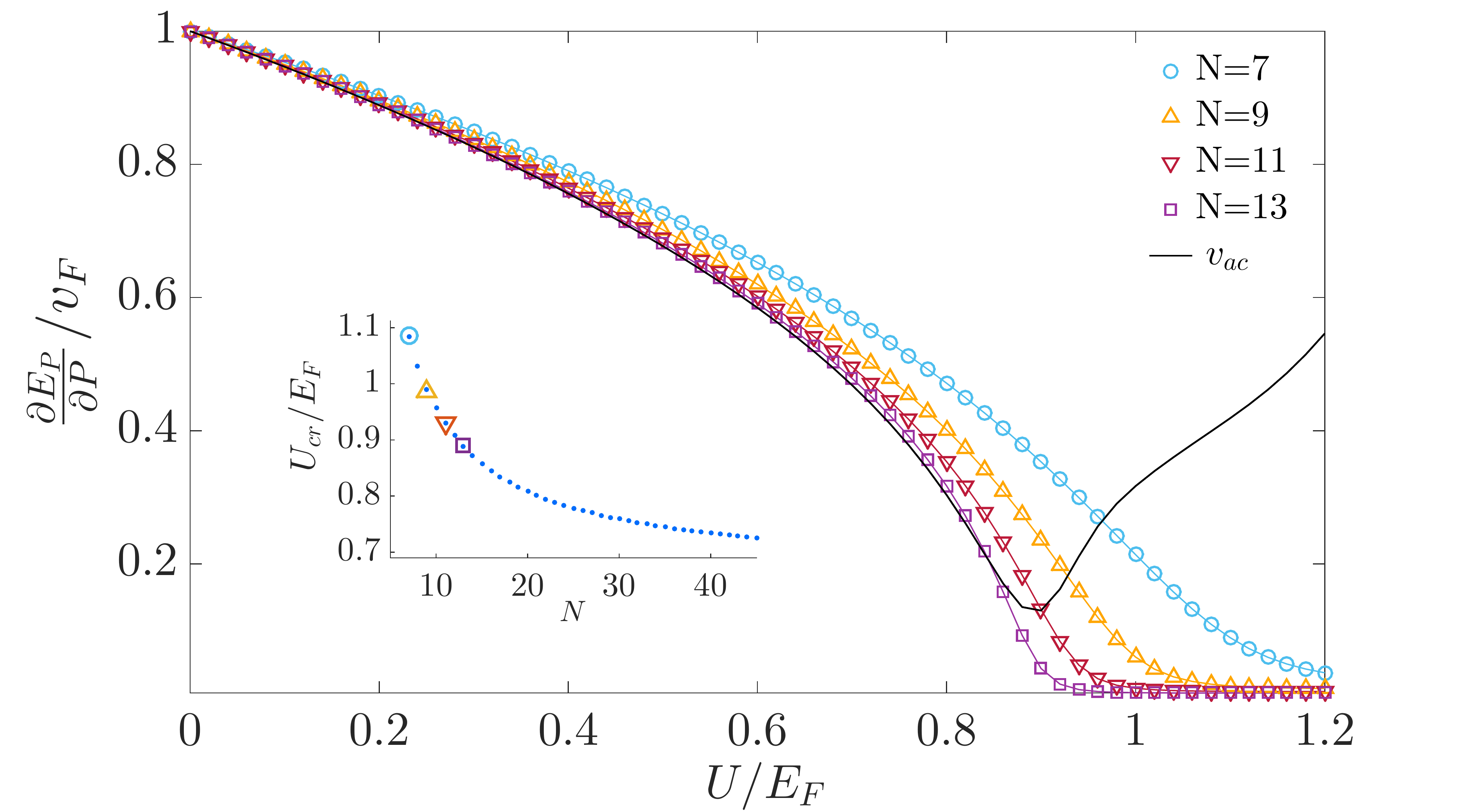}} b) \\
	\end{minipage}
	\caption{Quantum phase transition in a one-dimensional spinless fermionic system. a) The ground state energy $E_P$ as a function of the total momentum $P$ has a Luttinger liquid quasi-periodic arc shape for $U<U_{cr}$ and follows  the parabolic  dispersion law of a free particle with the mass $Nm$ for $U>U_{cr}$. The number of fermions is $N=11$; all curves are vertically shifted to start from the same origin;  $U$ in the legend is quoted in units of  $E_F$.  $\,$ b) Derivative $\frac{\partial E_P}{\partial P}|_{P\to0}$ as a function of interaction strength $U$ for a different number of fermions and fixed average density. The solid line represents a dependence of the acoustic mode velocity $v_{ac}$ for $N=13$, which starts to deviate from $\frac{\partial E_P}{\partial P}|_{P\to0}$ for $U>U_{cr}$.   Inset: the critical interaction $U_{cr}$ as a function of $N$ (color markers) and its fit $U_{cr}=\alpha_1+\alpha_2/N$ (blue dotted line).}
	\label{transition}
\end{figure}

One of the central objects of our study is the energy $E_P$ of the lowest eigenstate $|\Psi_P\rangle$ with a given total momentum $P$. The function $E_P$ is the lower edge of the many-body excitation spectrum.
Let us remind an important universal periodicity feature of $E_P$  \cite{haldane1981luttinger}. This feature follows from the fact that any two lowest eigenfunctions corresponding to momenta shifted by $ 2 k_F$ are related according to  $\Psi_{P+2k_F}(x_1, ..., x_N)=e^{ i \,\delta k \, \sum_j x_j } \Psi_P(x_1 , ..., x_N )$. The corresponding eigenenergies satisfy the relation
\begin{equation}\label{periodic}
E_P-\frac{P^2}{2 N m}=E_{P+ 2 k_F}-\frac{(P+ 2 k_F)^2}{2 N m}.
\end{equation}
This relation implies $E_P$ being a periodic function with the period $2 k_F$ for systems in the thermodynamical limit.

%

A typical family of $E_P$ obtained for different values of $U$ in our simulations is shown in Figure \ref{transition}a.  In accordance with eq.~(\ref{periodic}),   the points  $(P,E_P)$ for $P$ being multiples of $2 k_F$ lie on the parabola $\frac{P^2}{2 N m}$, i.e. on the dispersion curve of a single particle with the mass $N m$. In the case of weak interaction $E_P$ has a shape of periodic arcs, which is a signature of a Luttinger liquid \cite{haldane1981luttinger}.
This shape is qualitatively the same as for the free Fermi gas and for exactly-solvable models of bosons and fermions \cite{lieb1963exactI,gaudin1967systeme,yang1967some}. However,  for $U$ exceeding certain critical value $U_{cr}$, $E_P$ coincides with the parabola $\frac{P^2}{2 N m}$  for any $P$, that is the system behaves like a composite particle of the mass $N m$.  This manifests the phase transition to the phase-separated state, which occurs in the model (\ref{H}) as interaction strength $U$ increases. Quite an instructive qualitative picture of the phase separation in 1D quantum systems arises from the world-line interpretation, which we provide in Appendix along with several corollaries. \par

  \par

The phase transition can be characterized  using the velocity $\left.\frac{\partial E_P}{\partial P}\right|_{P=+0}$ estimated as $(E_{\delta k}- E_0)/\delta k$ for finite-size systems. 
Figure~\ref{transition}b shows the dependence of this quantity on $U$ for different $N$ and fixed average density $\rho\equiv N/L$. For small interaction  $\left.\frac{\partial E_P}{\partial P}\right|_{P=+0}$ equal to the velocity of a Luttinger liquid acoustic mode and therefore should be finite.
For $U$ above the critical value $\left.\frac{\partial E_P}{\partial P}\right|_{P=+0}$ goes to zero.
\begin{figure}[t]
	\begin{minipage}[h]{0.95\linewidth}
		\center{\includegraphics[width=\linewidth]{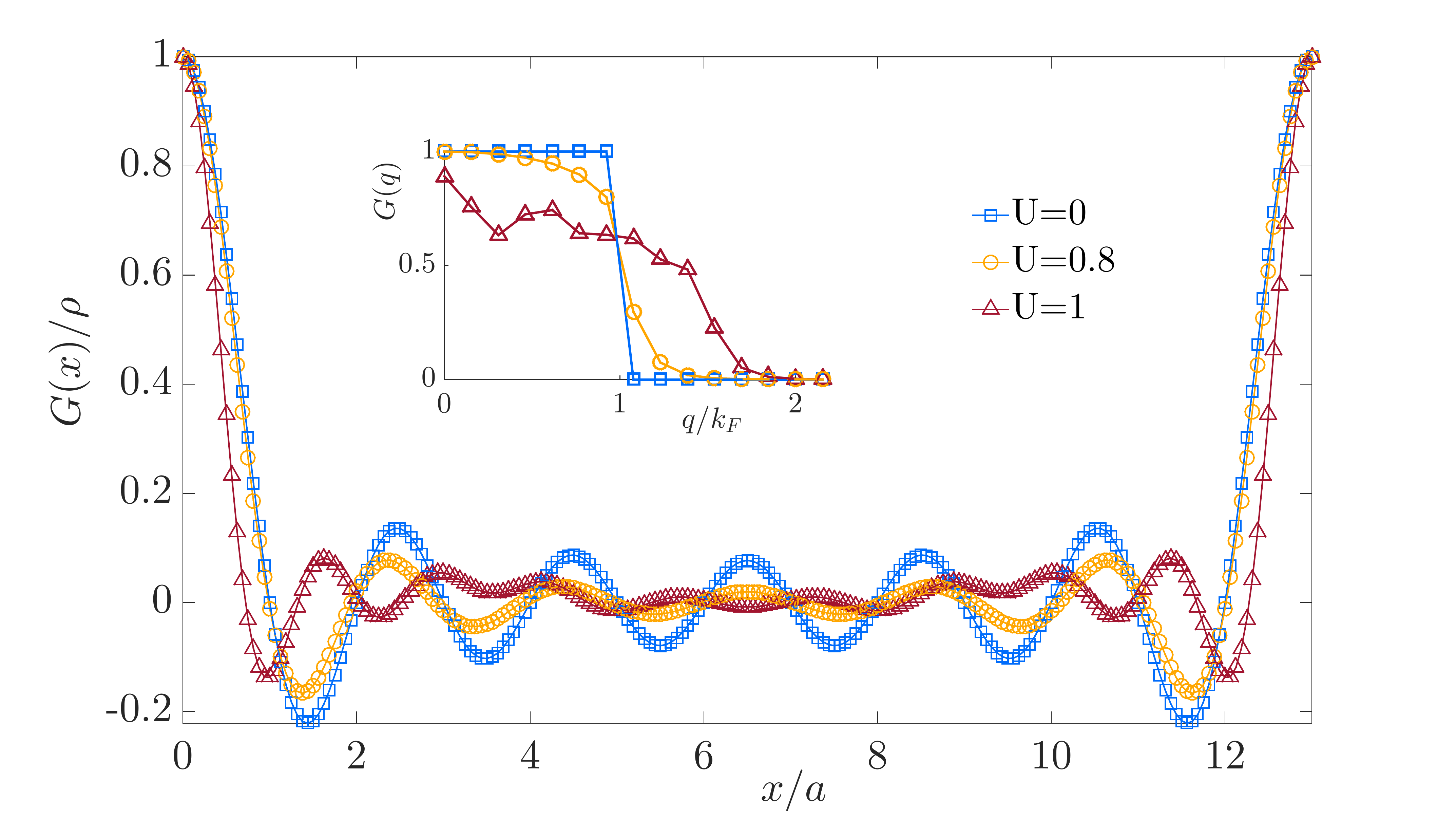}} a) \\
	\end{minipage}
	\vfill
	\begin{minipage}[h]{0.95\linewidth}
		\center{\includegraphics[width=\linewidth]{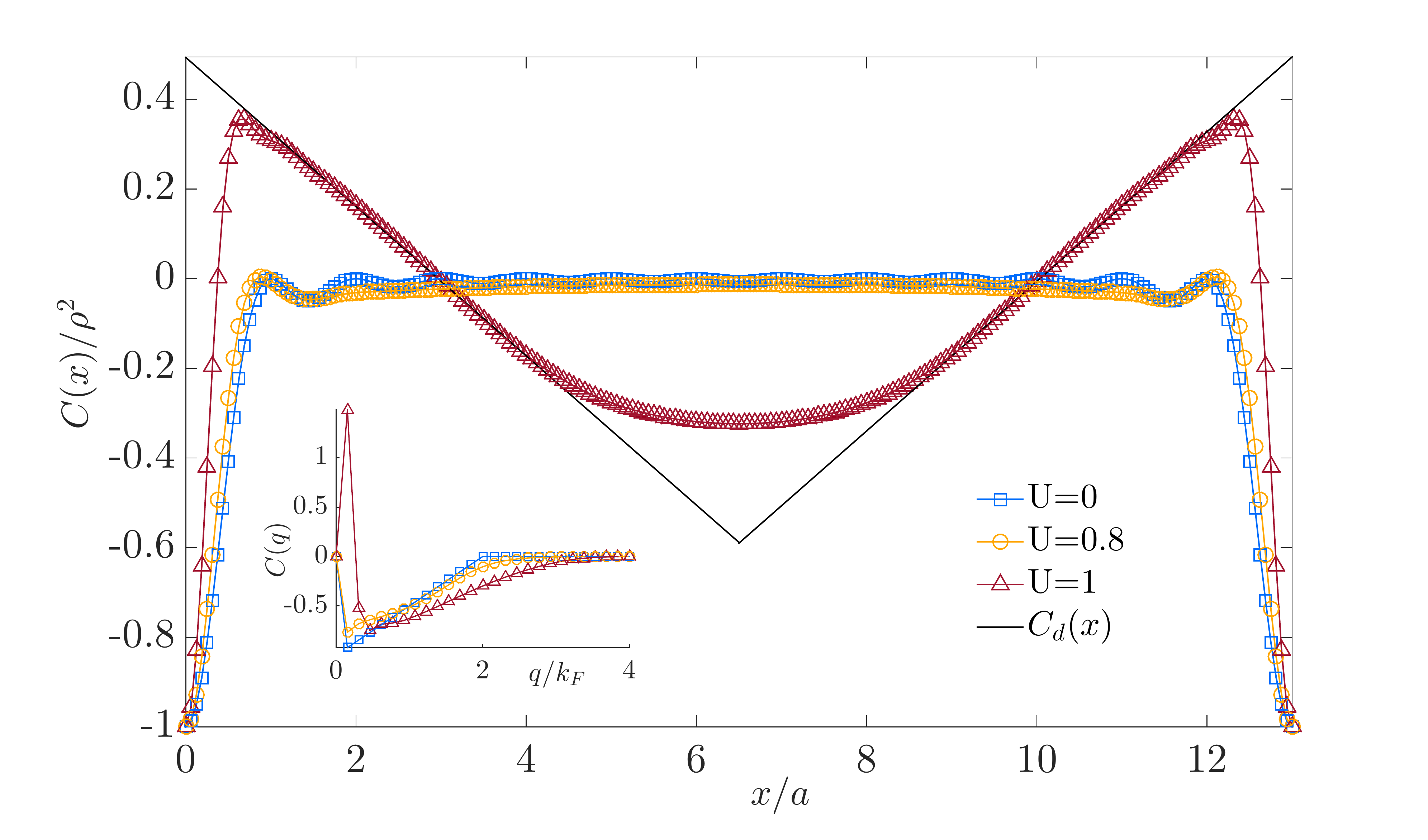}} b) \\
	\end{minipage}
	\caption{Correlation functions for free fermions (blue squares), the homogeneous phase $U<U_{cr}$ (yellow circles) and the droplet phase $U>U_{cr}$ (red triangles) for the state with zero total momentum and $N=13$ (the finite-size value of $U_{cr}$ for this $N$ is $0.89 E_F$). Interaction strength $U$ in the legends is quoted in units of  $E_F$. a) One-particle correlation function $G(x)=\langle \hcd_x\hc_{o}\rangle$. Its Fourier transform $G(q)$ is shown in the inset.  Observe the decrease of the oscillation period in $G(x)$  in the droplet phase.  $\,$ b) Density-density correlation function $C(x)=-\langle \hcd_x\hcd_o\hc_x\hc_o \rangle-\rho^2$. Its Fourier transform $C(q)$ is shown in the inset. Note the qualitative change in $C(x)$  and the emergence of a spike at $q=\delta k$ in $C(q)$ in the droplet phase. The black solid line represents a fit by the correlation function of a classical droplet (see Section \ref{subsec corfunc} for details). }
	\label{corf}
\end{figure}

\par
We have compared the above results with purely vibrational excitations of the system. Note that there is no lattice in the model (\ref{H}), and all excitations in this system are electronic ones. We considered the transition energy $E_{01}$ from the ground state to the first excited state at $P=0$. This excited state is composed of two excitations with the momentum $k=\pm \delta k$. Thus, assuming that the low lying collective modes are acoustic ones, i.e. have a linear spectrum, one estimates the  sound velocity as $v_{ac}=E_{01}/2 \delta k$. The latter quantity is plotted as a function of $U$ for $N=13$ in Figure \ref{transition}b.
One can see that below the critical interaction $v_{am}$ is indistinguishable from $\left.\frac{\partial E_P}{\partial P}\right|_{P=+0}$, which is consistent with the Luttinger liquid picture. At the transition point, the sound velocity falls to zero (in the thermodynamical limit), which means a diverging static compressibility. This is an evidence of the second-order transition to a phase-separated state above $U_{cr}$.
We use the minimum of $v_{ac}(U)$ for different values of $N$ to obtain the numerical estimates of the critical value $U_{cr}(N)$, which are shown as different color markers in the inset of Figure~\ref{transition}b. Finite-size scaling $U_{cr}(N)=\alpha_1+\alpha_2/N$ gives an estimation for the quantum critical point in the thermodynamical limit $U^{tl}_{cr}\equiv U_{cr}(N \to \infty) =\alpha_1=(0.65\pm 0.03) E_F$.



\subsection{Correlation functions and droplet density \label{subsec corfunc}}

 The phase transition in the model (\ref{H}) can also be observed in the correlation functions for the state with zero total momentum.  In Figure \ref{corf}a we show the one-particle correlation function $G(x)=\langle \hcd_x\hc_{o}\rangle$ for various values of interaction strength. Its Fourier transform $G(q)$ is shown in the inset. For free Fermi gas $G(x)$ is a periodized sinc function, i.e. Fourier transform of the step-function momentum distribution $G(q)$ (Fig. 2a, inset). Remind that the free Fermi gas has the Luttinger liquid parameter $K\!=\!1$\cite{Giamarchi2003}.
In the homogeneous phase, i.e. for $U<U_{cr}$, the correlation function $G(x)$ exhibits oscillations with the same period equal to $2a$, but they get damped faster. This displays a Luttinger liquid behavior with $K>1$ \cite{Giamarchi2003}, i.e. this parameter grows as $U\!\to \!U_{cr}$.  
 Note that the phase transition in the thermodynamical limit implies diverging $K$, but in our simulations for finite-size systems this parameter is always finite.
For $U>U_{cr}$ a remarkable change appears in the $G(x)$ -- its oscillations become more frequent, which signals the decrease of the distance between fermions, i.e. the  droplet formation. \par


Figure \ref{corf}b  shows the behavior of the density correlation $C(x)=-\langle \hcd_x\hcd_o\hc_x\hc_o \rangle-\rho^2$ during the phase transition. For the free Fermi gas $C(x)$ can be evaluated analytically as $C(x)=-|G(x)|^2$, by applying the Wick's theorem.  This gives a  squared sinc function (periodized due to the finite system size), i.e. Friedel oscillations. This picture remains qualitatively unchanged in the homogeneous phase, but oscillations damp as $U$ approaches to the transition point, where they are completely suppressed. Note that this behavior of the density correlation also implies the growth of the Luttinger parameter $K$ \cite{Giamarchi2003} with $U$, as long as $U<U_{cr}$. However in the droplet phase, i.e. for $U>U_{cr}$, both  $C(x)$ and its Fourier transform,  $C(q)$, change dramatically. Namely, $C(q)$ develops a spike at $q=\delta k$, which leads to $C(x)$ becomes similar to the correlation function of a classical droplet, $C_d(x)$. By the term 'classical droplet' we mean the system with uniformly distributed density over the 1D volume $L_d<L$, which we will refer as 'a droplet size'. Its correlation function can be defined as  $C_d(x_2-x_1) \equiv\langle \rho(0)\rho(x)\rangle_{cl}- \rho^2$, where $\langle\rho(0)\rho(x)\rangle_{cl}=\rho_d^2\, p(x_2-x_1)$ and $p(x)$ is the probability that a randomly chosen point $x_1$ and another point $(x_1+x)$ are both within a droplet of the size $L_d\equiv \rho L/\rho_d$. 
For $L_d\geq L/2$ one obtains $p(x)=\left(L_d-\min\{x,L-x\}\right)/L$ and  
\begin{equation}\label{Ccl}
C_d(x)=\begin{cases}\left(\rho/\rho_d -x/L\right) \rho_d^2- \rho^2, & |x|\le \frac{L}{2}
\\\left(\rho/\rho_d +x/L-1\right) \rho_d^2- \rho^2, & \frac{L}{2}<|x|\leq L\end{cases}
\end{equation}

\begin{figure}[t]
	\center{\includegraphics[width=0.95\linewidth]{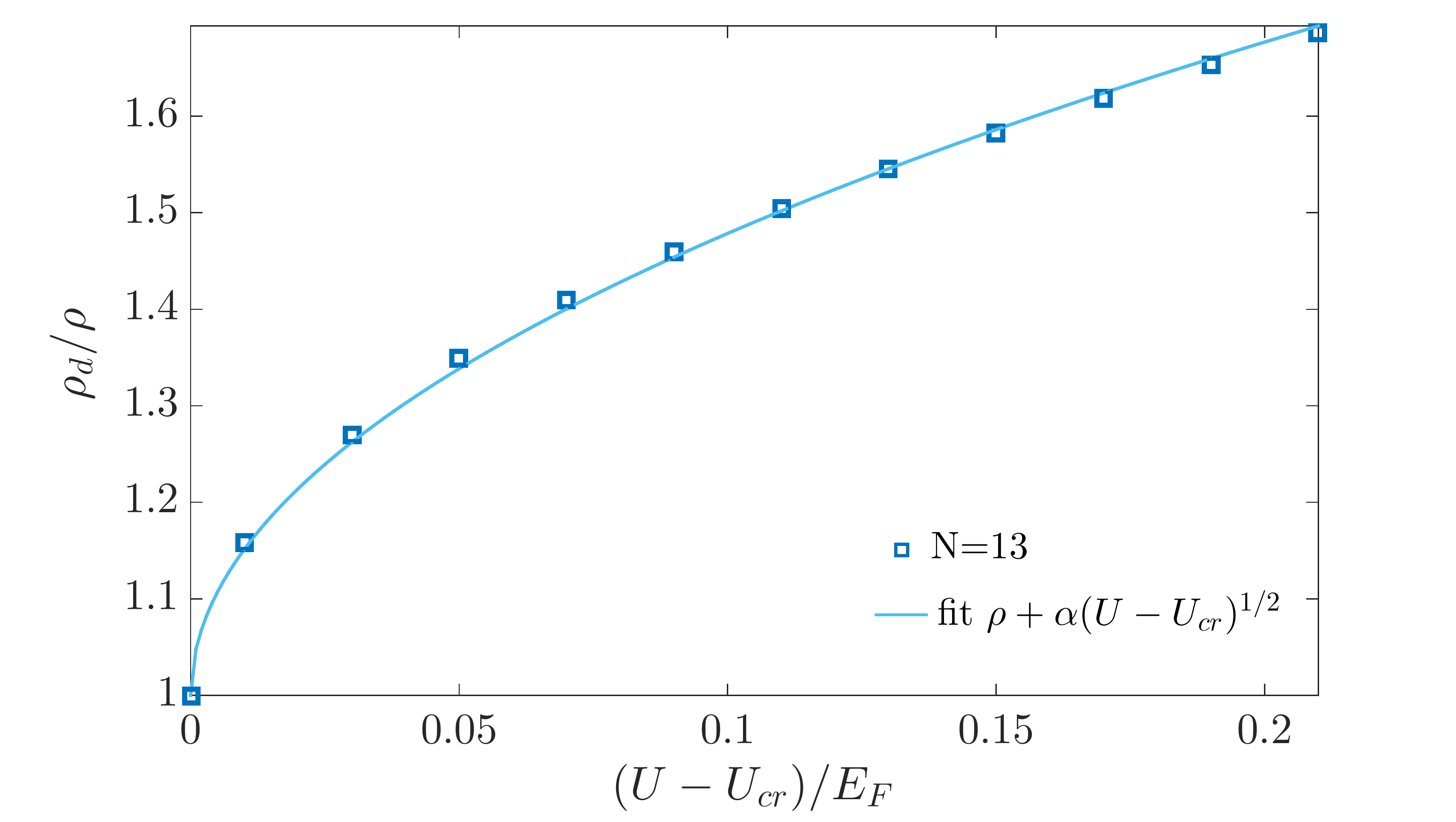}}
	\centering{}\caption{Density of the droplet $\rho_d$ as a function of interaction strength between particles. The number of particle is $N=13$, the finite-size value of $U_{cr}$ is $0.89 E_F$. The blue curve represents a fit $\rho_d=\rho+\alpha \sqrt{U-U_{cr}}$. }
	\label{density}
\end{figure}

We use this explicit formula to fit  $C(x)$  in the droplet phase (thin black line in Figure \ref{corf}b).
This fit allows us to estimate the droplet density from  the value of $C_d(x)$ at $x\!=\!0$ according to $\rho_d=\rho+C_d(x\!=\!0)/\rho$. We show thus obtained droplet densities $\rho_d$ for different interaction strengths $U$ in Figure \ref{density}.  The square-root dependence $\rho_d(U)$ seems to be determined by the quantum fluctuations of the droplet boundaries, which are well-known to be crucial for the critical region behavior \cite{sachdev2011quantum}. For larger $U$ unattainable in our numerical calculations this square-root dependence should crossover to the linear dependence implied by the mean-field estimate, which we discuss in the next subsection.

\subsection{Mean field estimate of the critical coupling}
It is instructive to compare the numerical result of exact diagonalization for the critical coupling with the mean field estimate.
The kinetic and potential energies of a fermionic droplet at rest in a free space  can be estimated as
\begin{align}
E_K= N\,\frac{\pi^2\rho_d^2}{6\,m},~~~~E_P=-N\,\rho_d\, U \,b,
\end{align}
where $\rho_d$ is the density of the droplet, and we assume that the average distance between the fermions in the droplet, $\rho_d^{-1}$, is on the order or smaller than the interaction range $b$.
Formally, the total energy, $E_K+E_P$, is minimized for $\rho_d = \frac{3}{\pi^2} b\,U\,m$.
However, for small enough $U$ the latter quantity is less than $\rho$. This means that for such $U$ the free droplet does not fit into the `container' of length $L$ and thus the homogeneous phase is favored as the ground state. Formation a bounded droplet with the density $\rho_d>\rho$  becomes energetically favorable (at the mean field level) for attraction exceeding
\begin{align}\label{critical interaction}
U_{mf}=\frac{\pi^2\rho}{3\,b\,m}
\end{align}
For our choice of the potential (\ref{Ur}) with $b=1.5\rho^{-1}$ the relation (\ref{critical interaction}) provides a mean-field estimate $U_{mf}= 0.45\, E_F$. Given the strongly interacting nature of the system at criticality, the agreement between $U_{mf}$ and $U^{tl}_{cr}$ is quite reasonable.

\subsection{Pressure}

\begin{figure}[b]
	\center{\includegraphics[width=0.99\linewidth]{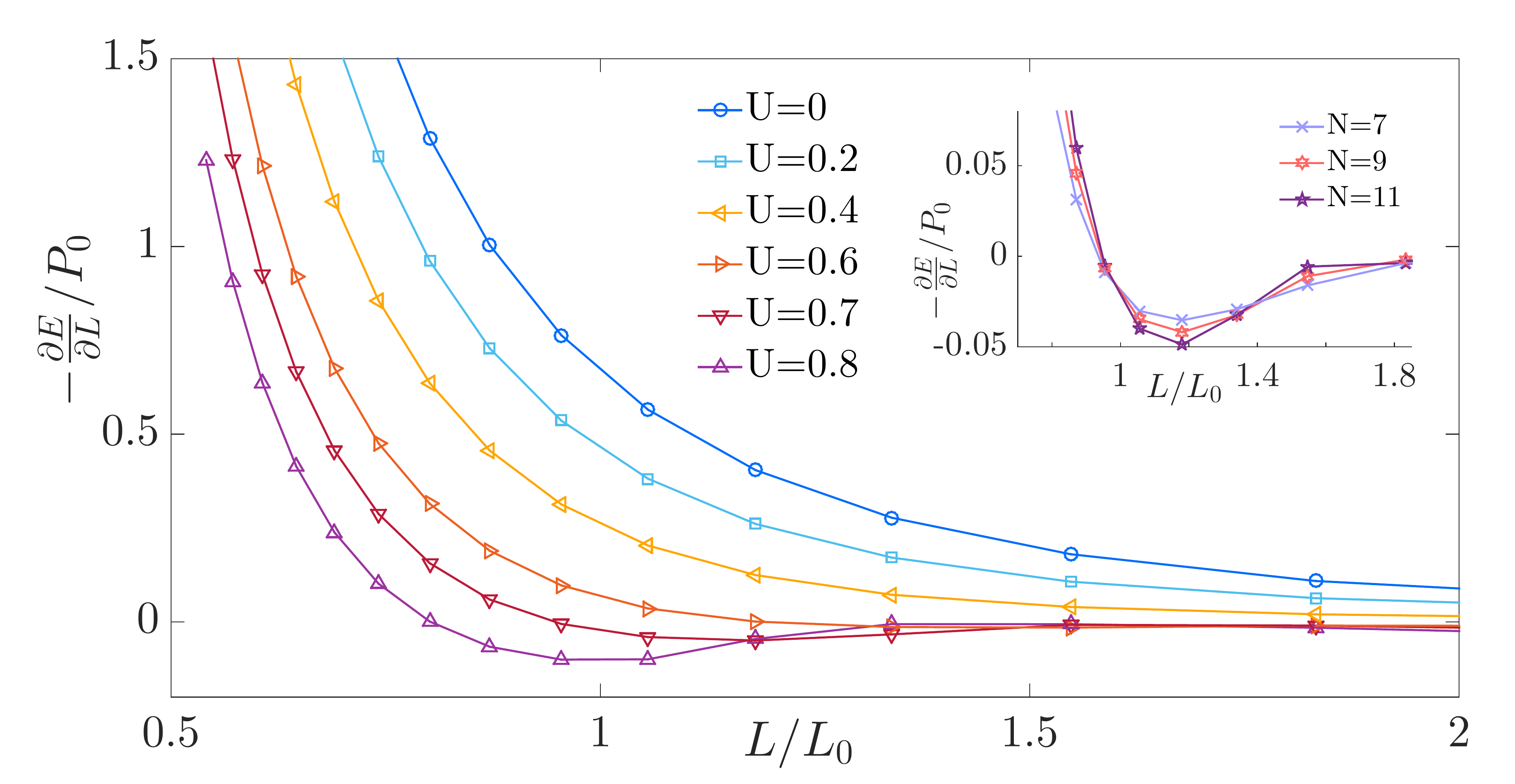}}
	\centering{}\caption{Pressure in a 1D fermionic system for different values of $U$ (in  units of  $E_F$ for a reference system size $L_0$) for $ N=11$. Inset: Zoomed view of a negative pressure region for $U=0.8E_F$ and a various number of particles $N$. The size of this region decreases with increasing $N$. }
	\label{pressure}
\end{figure}
\par Finally, we present the results on the  behavior of the pressure during the transition to the phase separated state. Since the pressure is determined by a variation of the energy with respect to the length of the system, in this subsection we fix a reference value $L_0$ of the length to be able to compare results for various values of $L$. $E_F$ and $k_F$ are  determined with respect to $L_0$. We also define a reference pressure $P_0\equiv k_F^3/2m$. 
	In Figure \ref{pressure} we show $P=-\frac{\partial E}{\partial L}$ for our model as a function of the system size.  In the homogeneous phase the pressure decreases as $L^{-3}$, analogously to the case of the free Fermi gas. With increasing $U$ the pressure goes to zero at some critical value $U_{cr}(L)$, and at this point particles form a localised droplet. This critical value weakly depends on the number of particles and can be used as an independent estimate of the critical coupling, $U_{cr}$, in the thermodynamical limit. For $L=L_0$ this estimate gives $\widetilde{U^{tl}_{cr}}=0.7 E_F$, which is close to the  value $U^{tl}_{cr}=0.65 E_F$ obtained in Section \ref{subsec spectral edge} by means of the finite-size scaling.

Quite remarkably, in the vicinity of the transition point there exists a small range of $L$ where the pressure is negative. We attribute this negative pressure to the effect of surface tension: in this regime the strength of attraction is just below the critical value necessary to overcome the surface tension and create the phase boundaries. In the inset to Figure \ref{pressure} we show how these negative pressure region shrinks as the number of fermions increases.

\par

\section{Discussion}
Let us make several remarks concerning the scope and possible extensions of our studies.
We have investigated the model (\ref{H})  for the specific choice of the interaction potential given by equation (\ref{Ur}). This choice has been mostly determined  by the limitations of exact diagonalization. However, we believe that the phase transition to the phase-separated state exists for a broad class of finite-range and effectively finite-range attractive potentials.
	Also, we suppose that it occurs even for arbitrary small  potential width, i.e. parameter $b$, but in this case, sufficiently large values of interacting strength $U$ are required. We have been able to check this statement in the regime when the parameter $b$  is close to the average interparticle distance $a$. As expected, the critical interaction $U_{cr}$ in our simulations becomes larger as the parameter $b$ decreases. Investigating the conditions for phase separation away from the regime $b\!\sim\! a$ is an intriguing problem, which may, however, require different theoretical and numerical approaches.




It is also instructive to compare the effect discussed in the present paper to a related effect which occurs in a system of attracting one-dimensional bosons. The latter system can be studied analytically by means of the Bethe ansatz provided the boson-boson interaction is point-like, as shown by Lieb and Lininger \cite{lieb1963exactI}. If this interaction is positive, the ground state is a bound state of all bosons \cite{lieb1963exactI, mcguire1964study}. However, this bosonic `supermolecule' has important differences from the fermionic droplet discussed here. First, it is not well defined in the thermodynamic limit -- particle density and energy per particle in the bosonic supermolecule diverge in this limit. Second, the supermolecule is formed at an arbitrarily small attraction, in contrast to a finite critical interaction strength in our case. \par

Last, let us briefly discuss the spinful fermions. In contrast to spinless fermions, already the point attraction  leads to the non-trivial physics in the case of spinful fermions, which can be reviled by the explicit Bethe ansatz solution \cite{gaudin1967systeme,yang1967some}. Qualitatively, the point attraction acting between fermions with opposite spin projections
leads to pairing which in the case of strong attraction can be sought of as singlet molecule formation \cite{hu2007phase,guan2007phase}. We note that pairing of fermions with coinciding spin projections, i.e. formation of triplet molecular states, is expected to be less energetically favorable.
In the molecular regime the spinful system can be described as a gas of hard-core bosons, which can be mapped on the free Fermi gas  in one dimension \cite{Girardeau1960}.   If now a nonlocal attraction between the  spinful fermions  is introduced  in addition to the point-like attraction, one can expect that it  will cause the finite-range attraction between these singlet molecules. Since there is a mapping of hard-core bosons to the model (\ref{H}), sufficiently large interaction should eventually lead  to the droplet formation.
We believe that studying the interplay between the pairing and the droplet formation in a  spinful fermion system is an interesting avenue for the future research. 	 
\section{Conclusions}
 In the present work, we have investigated the droplet formation in the 1D spinless fermionic system with a finite-range attraction.
Although many scenarios of the phase transitions are forbidden in translationally invariant 1D systems due to quantum fluctuations, a quantum phase transition to a phase-separated state was previously predicted in the considered system. In the present work we have verified the existence of the phase separation by the exact diagonalization study and performed the finite-size scaling to estimate the value of the critical interaction. We have demonstrated that the phase-separated state
apparently lacks the conventional Luttinger liquid features.
In particular, the center-of-mass motion becomes independent from the vibrational degrees of freedom and the system is characterized by a parabolic dispersion law, in sharp contrast to the Luttinger liquid behavior. In fact, in the phase-separated state, the system can be described as a droplet of a liquid in the thermodynamical sense. We have studied the details of the droplet formation and its properties by numerically calculating the pressure as a function of the interaction strength as well as the correlation functions.
The critical interaction strength in the thermodynamic limit has been estimated by two complementary methods with consistent results. We have also obtained a square-root dependence of the droplet density over the interaction strength in the vicinity of the critical point. Our results are also applicable for the case of hard-core bosons with finite-range attraction. We believe that our work can inspire further theoretical and experimental investigation of such systems. 


\subsection*{ Acknowledgements}
We are grateful to V. V. Cheianov and V. I. Yudson  for a fruitful discussion and valuable remarks.
This work  was funded by the RSF Grant No. 16-42-01057

\subsection*{Appendix: World-line description of phase separation}
\begin{figure}[t]
	\center{\includegraphics[width=0.75\linewidth]{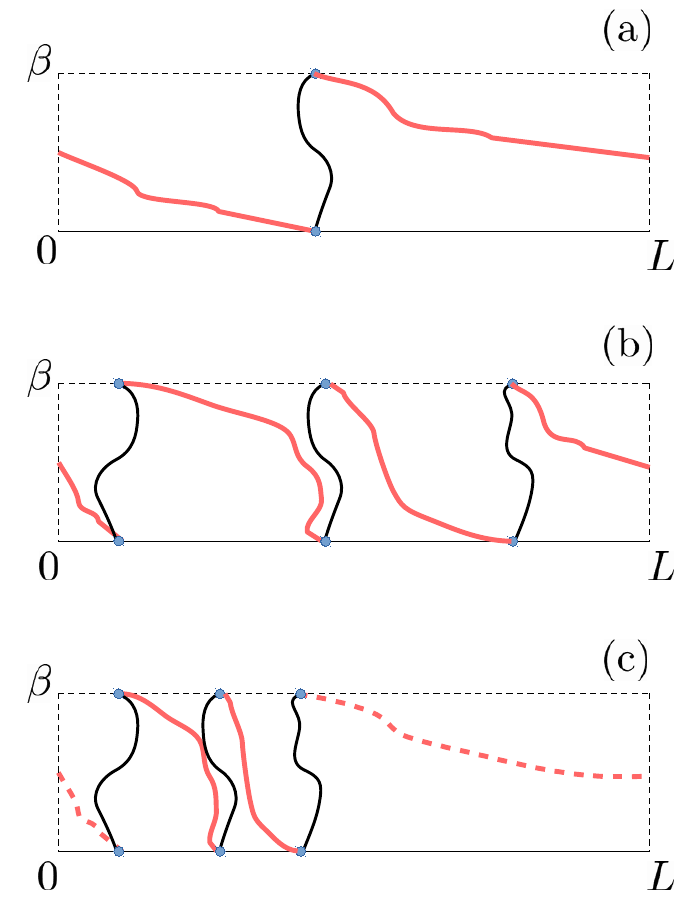}}
	\centering{}\caption{Imaginary-time Feynman trajectories of the particles moving in a periodized system. $\,$a) For a single particle the trajectory  returns to an initial point as evolving from to $\tau=0$ to $\beta$ (thin black line) or performs a shift by an integer number of the quantization box lengths $L$ (thick red line shows  a shift by $L$).  $\,$ b) For several indistinguishable particles, there are cyclic  shifts. $\,$ c) For the phase separation, cyclic shifts are suppressed, as one of the particles should leave the drop during the evolution (dashed red line).}
	\label{figWL}
\end{figure}

Let us start from the analysis of  the collective dispersion law in a periodized 1D system.
It is useful to consider possible imaginary-time Feynman trajectories of fermions, shown in Figure \ref{figWL}. For a single particle evolving in an infinite space, the trajectory is periodic in imaginary time $\tau$, that is $x(\tau=0)=x(\tau=\beta)$ (for the zero temperature case we consider,  $\beta$ should be formally taken infinitely large). As panel (1) of Figure \ref{figWL} illustrates, periodization introduces another type of trajectories, where the particle returns to a periodized copy of itself. It means that the coordinate shifts by a number of the quantization periods, $x(\tau=0)=x(\tau=\beta)+l L$ with an integer $l$.  It can be shown that a presence of these trajectories gives rise to the momentum quantization $k=l \delta k$ for a single particle.

Let us now consider trajectories for $N$ identical particles, starting at  some $x_j(\tau=0)$. In higher dimensions there are exchange processes, so that in an infinite space the values of $x_j(\tau=0)$ and $x_j(\tau=\beta)$ coincide but can be reordered within the set. However the Pauli principle forbids crossing of the trajectories for fermionic particles, thus making exchange processes impossible in infinite 1D space.  For a periodized system,
the only possible exchange process is a permutation of all particles in a circle, $x_j(\tau=0)=x_{j+1}(\tau=\beta)$ modulo $L$. An example of such trajectories is given in the panel (b) of Figure \ref{figWL}.
Considering for simplicity an equidistant initial placement of the particles, one sees that each particle gets shifted  by the interparticle distance,
$x_j(\tau=0)=x_{j}(\tau=\beta)+l L/N$. This suggests the  $2 k_F$ periodicity of the collective dispersion law in a thermodynamical limit.

We make an important observation that the $2 k_F$ periodicity of the collective dispersion law
are inevitably related to the quantum indistinguishability. In 1D, identical fermions are very similar to any particles with a point hard core repulsion (Paulions).
No matter, are these particles identical or not, hard core interaction forbids the trajectory crossings thus virtually imposing the Pauli principle.
Consequently e.g. spectra of acoustic modes of fermions and distinguishable hard core particles
coincide, provided the same masses and non-local part of the interaction potential. However,  permutations in a circle are possible for identical particles only.  Distinguishability will only allow for all particles being simultaneously shifted by $l L$.
In terms of wave functions  of non-identical particles, a transformation $\Psi(x_1, ..., x_j, ..., x_N) \to  e^{i\delta k x_j}\Psi(x_1, ...,  x_j, ..., x_N)$ can be considered
yielding $E_P=\frac{P^2}{2 N m}$ with $P=l \delta k$.

Processes related to circular permutations are relevant only when particles are placed nearly equidistantly. Panel (c) of Figure \ref{figWL} shows a situation of the phase separation, when particles form a localized droplet. In this case a permutation requires one particle leaving the droplet and moving fast to reach another side of the drop,  as the dashed line indicates. The larger the system is, the less probable such processes are. Therefore phase separation   suppresses the exchange processes and thus leads to a parabolic dispersion.

\par
Our study provides also an understanding why the phase separation is not destroyed by the quantum fluctuations. Indeed, in the phase separated regime, the droplet's center of mass degree of freedom is decoupled from vibration modes. Therefore
zero-point vibrations do not affect the droplet average position, and Mermin-Wagner argumentation does not apply.
In the thermodynamical limit of a periodized  system, an order parameter can be introduced: one can take e.g. the average phase $2 \pi \av{x}/L$. This is in a very contrast with a uniform Luttinger liquid, where
the collective degree of freedom  is  intrinsically connected to the acoustic excitations (e.g. this is why $\left.\frac{\partial E_P}{\partial P}\right|_{P=+0}$ coincides with the sound velocity in this regime).


\bibliography{1D}

\providecommand{\noopsort}[1]{}\providecommand{\singleletter}[1]{#1}%
\begin{thebibliography}{36}%
\makeatletter
\providecommand \@ifxundefined [1]{%
 \@ifx{#1\undefined}
}%
\providecommand \@ifnum [1]{%
 \ifnum #1\expandafter \@firstoftwo
 \else \expandafter \@secondoftwo
 \fi
}%
\providecommand \@ifx [1]{%
 \ifx #1\expandafter \@firstoftwo
 \else \expandafter \@secondoftwo
 \fi
}%
\providecommand \natexlab [1]{#1}%
\providecommand \enquote  [1]{``#1''}%
\providecommand \bibnamefont  [1]{#1}%
\providecommand \bibfnamefont [1]{#1}%
\providecommand \citenamefont [1]{#1}%
\providecommand \href@noop [0]{\@secondoftwo}%
\providecommand \href [0]{\begingroup \@sanitize@url \@href}%
\providecommand \@href[1]{\@@startlink{#1}\@@href}%
\providecommand \@@href[1]{\endgroup#1\@@endlink}%
\providecommand \@sanitize@url [0]{\catcode `\\12\catcode `\$12\catcode
  `\&12\catcode `\#12\catcode `\^12\catcode `\_12\catcode `\%12\relax}%
\providecommand \@@startlink[1]{}%
\providecommand \@@endlink[0]{}%
\providecommand \url  [0]{\begingroup\@sanitize@url \@url }%
\providecommand \@url [1]{\endgroup\@href {#1}{\urlprefix }}%
\providecommand \urlprefix  [0]{URL }%
\providecommand \Eprint [0]{\href }%
\providecommand \doibase [0]{http://dx.doi.org/}%
\providecommand \selectlanguage [0]{\@gobble}%
\providecommand \bibinfo  [0]{\@secondoftwo}%
\providecommand \bibfield  [0]{\@secondoftwo}%
\providecommand \translation [1]{[#1]}%
\providecommand \BibitemOpen [0]{}%
\providecommand \bibitemStop [0]{}%
\providecommand \bibitemNoStop [0]{.\EOS\space}%
\providecommand \EOS [0]{\spacefactor3000\relax}%
\providecommand \BibitemShut  [1]{\csname bibitem#1\endcsname}%
\let\auto@bib@innerbib\@empty
\bibitem [{\citenamefont {Volovik}(2003)}]{volovik2003universe}%
  \BibitemOpen
  \bibfield  {author} {\bibinfo {author} {\bibfnamefont {G.~E.}\ \bibnamefont
  {Volovik}},\ }\href@noop {} {\emph {\bibinfo {title} {The universe in a
  helium droplet}}},\ Vol.\ \bibinfo {volume} {117}\ (\bibinfo  {publisher}
  {Oxford University Press on Demand},\ \bibinfo {year} {2003})\BibitemShut
  {NoStop}%
\bibitem [{\citenamefont {Toennies}\ and\ \citenamefont
  {Vilesov}(2004)}]{toennies2004superfluid}%
  \BibitemOpen
  \bibfield  {author} {\bibinfo {author} {\bibfnamefont {J.~P.}\ \bibnamefont
  {Toennies}}\ and\ \bibinfo {author} {\bibfnamefont {A.~F.}\ \bibnamefont
  {Vilesov}},\ }\href@noop {} {\bibfield  {journal} {\bibinfo  {journal}
  {Angewandte Chemie International Edition}\ }\textbf {\bibinfo {volume}
  {43}},\ \bibinfo {pages} {2622} (\bibinfo {year} {2004})}\BibitemShut
  {NoStop}%
\bibitem [{\citenamefont {Schmitt}\ \emph {et~al.}(2016)\citenamefont
  {Schmitt}, \citenamefont {Wenzel}, \citenamefont {B{\"o}ttcher},
  \citenamefont {Ferrier-Barbut},\ and\ \citenamefont
  {Pfau}}]{schmitt2016self}%
  \BibitemOpen
  \bibfield  {author} {\bibinfo {author} {\bibfnamefont {M.}~\bibnamefont
  {Schmitt}}, \bibinfo {author} {\bibfnamefont {M.}~\bibnamefont {Wenzel}},
  \bibinfo {author} {\bibfnamefont {F.}~\bibnamefont {B{\"o}ttcher}}, \bibinfo
  {author} {\bibfnamefont {I.}~\bibnamefont {Ferrier-Barbut}}, \ and\ \bibinfo
  {author} {\bibfnamefont {T.}~\bibnamefont {Pfau}},\ }\href@noop {} {\bibfield
   {journal} {\bibinfo  {journal} {Nature}\ }\textbf {\bibinfo {volume}
  {539}},\ \bibinfo {pages} {259} (\bibinfo {year} {2016})}\BibitemShut
  {NoStop}%
\bibitem [{\citenamefont {Ferrier-Barbut}\ \emph {et~al.}(2016)\citenamefont
  {Ferrier-Barbut}, \citenamefont {Kadau}, \citenamefont {Schmitt},
  \citenamefont {Wenzel},\ and\ \citenamefont
  {Pfau}}]{ferrier-barbut2016observation}%
  \BibitemOpen
  \bibfield  {author} {\bibinfo {author} {\bibfnamefont {I.}~\bibnamefont
  {Ferrier-Barbut}}, \bibinfo {author} {\bibfnamefont {H.}~\bibnamefont
  {Kadau}}, \bibinfo {author} {\bibfnamefont {M.}~\bibnamefont {Schmitt}},
  \bibinfo {author} {\bibfnamefont {M.}~\bibnamefont {Wenzel}}, \ and\ \bibinfo
  {author} {\bibfnamefont {T.}~\bibnamefont {Pfau}},\ }\href {\doibase
  10.1103/PhysRevLett.116.215301} {\bibfield  {journal} {\bibinfo  {journal}
  {Phys. Rev. Lett.}\ }\textbf {\bibinfo {volume} {116}},\ \bibinfo {pages}
  {215301} (\bibinfo {year} {2016})}\BibitemShut {NoStop}%
\bibitem [{\citenamefont {Cheiney}\ \emph {et~al.}(2018)\citenamefont
  {Cheiney}, \citenamefont {Cabrera}, \citenamefont {Sanz}, \citenamefont
  {Naylor}, \citenamefont {Tanzi},\ and\ \citenamefont
  {Tarruell}}]{cheiney2018bright}%
  \BibitemOpen
  \bibfield  {author} {\bibinfo {author} {\bibfnamefont {P.}~\bibnamefont
  {Cheiney}}, \bibinfo {author} {\bibfnamefont {C.~R.}\ \bibnamefont
  {Cabrera}}, \bibinfo {author} {\bibfnamefont {J.}~\bibnamefont {Sanz}},
  \bibinfo {author} {\bibfnamefont {B.}~\bibnamefont {Naylor}}, \bibinfo
  {author} {\bibfnamefont {L.}~\bibnamefont {Tanzi}}, \ and\ \bibinfo {author}
  {\bibfnamefont {L.}~\bibnamefont {Tarruell}},\ }\href {\doibase
  10.1103/PhysRevLett.120.135301} {\bibfield  {journal} {\bibinfo  {journal}
  {Phys. Rev. Lett.}\ }\textbf {\bibinfo {volume} {120}},\ \bibinfo {pages}
  {135301} (\bibinfo {year} {2018})}\BibitemShut {NoStop}%
\bibitem [{\citenamefont {Cabrera}\ \emph {et~al.}(2018)\citenamefont
  {Cabrera}, \citenamefont {Tanzi}, \citenamefont {Sanz}, \citenamefont
  {Naylor}, \citenamefont {Thomas}, \citenamefont {Cheiney},\ and\
  \citenamefont {Tarruell}}]{cabrera2018quantum}%
  \BibitemOpen
  \bibfield  {author} {\bibinfo {author} {\bibfnamefont {C.}~\bibnamefont
  {Cabrera}}, \bibinfo {author} {\bibfnamefont {L.}~\bibnamefont {Tanzi}},
  \bibinfo {author} {\bibfnamefont {J.}~\bibnamefont {Sanz}}, \bibinfo {author}
  {\bibfnamefont {B.}~\bibnamefont {Naylor}}, \bibinfo {author} {\bibfnamefont
  {P.}~\bibnamefont {Thomas}}, \bibinfo {author} {\bibfnamefont
  {P.}~\bibnamefont {Cheiney}}, \ and\ \bibinfo {author} {\bibfnamefont
  {L.}~\bibnamefont {Tarruell}},\ }\href@noop {} {\bibfield  {journal}
  {\bibinfo  {journal} {Science}\ }\textbf {\bibinfo {volume} {359}},\ \bibinfo
  {pages} {301} (\bibinfo {year} {2018})}\BibitemShut {NoStop}%
\bibitem [{\citenamefont {Bulgac}(2002)}]{bulgac2002dilute}%
  \BibitemOpen
  \bibfield  {author} {\bibinfo {author} {\bibfnamefont {A.}~\bibnamefont
  {Bulgac}},\ }\href {\doibase 10.1103/PhysRevLett.89.050402} {\bibfield
  {journal} {\bibinfo  {journal} {Phys. Rev. Lett.}\ }\textbf {\bibinfo
  {volume} {89}},\ \bibinfo {pages} {050402} (\bibinfo {year}
  {2002})}\BibitemShut {NoStop}%
\bibitem [{\citenamefont {Petrov}(2015)}]{petrov2015quantum}%
  \BibitemOpen
  \bibfield  {author} {\bibinfo {author} {\bibfnamefont {D.~S.}\ \bibnamefont
  {Petrov}},\ }\href {\doibase 10.1103/PhysRevLett.115.155302} {\bibfield
  {journal} {\bibinfo  {journal} {Phys. Rev. Lett.}\ }\textbf {\bibinfo
  {volume} {115}},\ \bibinfo {pages} {155302} (\bibinfo {year}
  {2015})}\BibitemShut {NoStop}%
\bibitem [{\citenamefont {W\"achtler}\ and\ \citenamefont
  {Santos}(2016)}]{wachtler2016ground-state}%
  \BibitemOpen
  \bibfield  {author} {\bibinfo {author} {\bibfnamefont {F.}~\bibnamefont
  {W\"achtler}}\ and\ \bibinfo {author} {\bibfnamefont {L.}~\bibnamefont
  {Santos}},\ }\href {\doibase 10.1103/PhysRevA.94.043618} {\bibfield
  {journal} {\bibinfo  {journal} {Phys. Rev. A}\ }\textbf {\bibinfo {volume}
  {94}},\ \bibinfo {pages} {043618} (\bibinfo {year} {2016})}\BibitemShut
  {NoStop}%
\bibitem [{\citenamefont {Macia}\ \emph {et~al.}(2016)\citenamefont {Macia},
  \citenamefont {S\'anchez-Baena}, \citenamefont {Boronat},\ and\ \citenamefont
  {Mazzanti}}]{macia2016droplets}%
  \BibitemOpen
  \bibfield  {author} {\bibinfo {author} {\bibfnamefont {A.}~\bibnamefont
  {Macia}}, \bibinfo {author} {\bibfnamefont {J.}~\bibnamefont
  {S\'anchez-Baena}}, \bibinfo {author} {\bibfnamefont {J.}~\bibnamefont
  {Boronat}}, \ and\ \bibinfo {author} {\bibfnamefont {F.}~\bibnamefont
  {Mazzanti}},\ }\href {\doibase 10.1103/PhysRevLett.117.205301} {\bibfield
  {journal} {\bibinfo  {journal} {Phys. Rev. Lett.}\ }\textbf {\bibinfo
  {volume} {117}},\ \bibinfo {pages} {205301} (\bibinfo {year}
  {2016})}\BibitemShut {NoStop}%
\bibitem [{\citenamefont {Sekino}\ and\ \citenamefont
  {Nishida}(2018)}]{sekino2018quantum}%
  \BibitemOpen
  \bibfield  {author} {\bibinfo {author} {\bibfnamefont {Y.}~\bibnamefont
  {Sekino}}\ and\ \bibinfo {author} {\bibfnamefont {Y.}~\bibnamefont
  {Nishida}},\ }\href {\doibase 10.1103/PhysRevA.97.011602} {\bibfield
  {journal} {\bibinfo  {journal} {Phys. Rev. A}\ }\textbf {\bibinfo {volume}
  {97}},\ \bibinfo {pages} {011602} (\bibinfo {year} {2018})}\BibitemShut
  {NoStop}%
\bibitem [{\citenamefont {Astrakharchik}\ and\ \citenamefont
  {Malomed}(2018)}]{astrakharchik2018dynamics}%
  \BibitemOpen
  \bibfield  {author} {\bibinfo {author} {\bibfnamefont {G.}~\bibnamefont
  {Astrakharchik}}\ and\ \bibinfo {author} {\bibfnamefont {B.}~\bibnamefont
  {Malomed}},\ }\href@noop {} {\bibfield  {journal} {\bibinfo  {journal} {arXiv
  preprint arXiv:1803.07165}\ } (\bibinfo {year} {2018})}\BibitemShut {NoStop}%
\bibitem [{\citenamefont {Nakamura}\ and\ \citenamefont
  {Nomura}(1997)}]{nakamura1997phase}%
  \BibitemOpen
  \bibfield  {author} {\bibinfo {author} {\bibfnamefont {M.}~\bibnamefont
  {Nakamura}}\ and\ \bibinfo {author} {\bibfnamefont {K.}~\bibnamefont
  {Nomura}},\ }\href {\doibase 10.1103/PhysRevB.56.12840} {\bibfield  {journal}
  {\bibinfo  {journal} {Phys. Rev. B}\ }\textbf {\bibinfo {volume} {56}},\
  \bibinfo {pages} {12840} (\bibinfo {year} {1997})}\BibitemShut {NoStop}%
\bibitem [{\citenamefont {Cabra}\ and\ \citenamefont
  {Drut}(2003)}]{cabra2003instabilities}%
  \BibitemOpen
  \bibfield  {author} {\bibinfo {author} {\bibfnamefont {D.}~\bibnamefont
  {Cabra}}\ and\ \bibinfo {author} {\bibfnamefont {J.}~\bibnamefont {Drut}},\
  }\href@noop {} {\bibfield  {journal} {\bibinfo  {journal} {Journal of
  Physics: Condensed Matter}\ }\textbf {\bibinfo {volume} {15}},\ \bibinfo
  {pages} {1445} (\bibinfo {year} {2003})}\BibitemShut {NoStop}%
\bibitem [{\citenamefont {Kolomeisky}\ \emph {et~al.}(2003)\citenamefont
  {Kolomeisky}, \citenamefont {Qi},\ and\ \citenamefont
  {Timmins}}]{kolomeisky2003ground-state}%
  \BibitemOpen
  \bibfield  {author} {\bibinfo {author} {\bibfnamefont {E.~B.}\ \bibnamefont
  {Kolomeisky}}, \bibinfo {author} {\bibfnamefont {X.}~\bibnamefont {Qi}}, \
  and\ \bibinfo {author} {\bibfnamefont {M.}~\bibnamefont {Timmins}},\
  }\href@noop {} {\bibfield  {journal} {\bibinfo  {journal} {Physical Review
  B}\ }\textbf {\bibinfo {volume} {67}},\ \bibinfo {pages} {165407} (\bibinfo
  {year} {2003})}\BibitemShut {NoStop}%
\bibitem [{\citenamefont {Kolomeisky}\ and\ \citenamefont
  {Timmins}(2005)}]{kolomeisky2005quantum}%
  \BibitemOpen
  \bibfield  {author} {\bibinfo {author} {\bibfnamefont {E.~B.}\ \bibnamefont
  {Kolomeisky}}\ and\ \bibinfo {author} {\bibfnamefont {M.}~\bibnamefont
  {Timmins}},\ }\href {\doibase 10.1103/PhysRevLett.95.226103} {\bibfield
  {journal} {\bibinfo  {journal} {Phys. Rev. Lett.}\ }\textbf {\bibinfo
  {volume} {95}},\ \bibinfo {pages} {226103} (\bibinfo {year}
  {2005})}\BibitemShut {NoStop}%
\bibitem [{\citenamefont {Law}\ and\ \citenamefont
  {Feldman}(2008)}]{law2008quantum}%
  \BibitemOpen
  \bibfield  {author} {\bibinfo {author} {\bibfnamefont {K.~T.}\ \bibnamefont
  {Law}}\ and\ \bibinfo {author} {\bibfnamefont {D.~E.}\ \bibnamefont
  {Feldman}},\ }\href {\doibase 10.1103/PhysRevLett.101.096401} {\bibfield
  {journal} {\bibinfo  {journal} {Phys. Rev. Lett.}\ }\textbf {\bibinfo
  {volume} {101}},\ \bibinfo {pages} {096401} (\bibinfo {year}
  {2008})}\BibitemShut {NoStop}%
\bibitem [{\citenamefont {Z\"ollner}(2011)}]{zolner2011ground}%
  \BibitemOpen
  \bibfield  {author} {\bibinfo {author} {\bibfnamefont {S.}~\bibnamefont
  {Z\"ollner}},\ }\href {\doibase 10.1103/PhysRevA.84.063619} {\bibfield
  {journal} {\bibinfo  {journal} {Phys. Rev. A}\ }\textbf {\bibinfo {volume}
  {84}},\ \bibinfo {pages} {063619} (\bibinfo {year} {2011})}\BibitemShut
  {NoStop}%
\bibitem [{\citenamefont {Yan}\ \emph {et~al.}(2014)\citenamefont {Yan},
  \citenamefont {Chen},\ and\ \citenamefont {Wan}}]{yan2014long}%
  \BibitemOpen
  \bibfield  {author} {\bibinfo {author} {\bibfnamefont {Z.}~\bibnamefont
  {Yan}}, \bibinfo {author} {\bibfnamefont {L.}~\bibnamefont {Chen}}, \ and\
  \bibinfo {author} {\bibfnamefont {S.}~\bibnamefont {Wan}},\ }\href@noop {}
  {\bibfield  {journal} {\bibinfo  {journal} {Journal of Physics B: Atomic,
  Molecular and Optical Physics}\ }\textbf {\bibinfo {volume} {47}},\ \bibinfo
  {pages} {135302} (\bibinfo {year} {2014})}\BibitemShut {NoStop}%
\bibitem [{\citenamefont {Edler}\ \emph {et~al.}(2017)\citenamefont {Edler},
  \citenamefont {Mishra}, \citenamefont {W\"achtler}, \citenamefont {Nath},
  \citenamefont {Sinha},\ and\ \citenamefont {Santos}}]{edler2015quantum}%
  \BibitemOpen
  \bibfield  {author} {\bibinfo {author} {\bibfnamefont {D.}~\bibnamefont
  {Edler}}, \bibinfo {author} {\bibfnamefont {C.}~\bibnamefont {Mishra}},
  \bibinfo {author} {\bibfnamefont {F.}~\bibnamefont {W\"achtler}}, \bibinfo
  {author} {\bibfnamefont {R.}~\bibnamefont {Nath}}, \bibinfo {author}
  {\bibfnamefont {S.}~\bibnamefont {Sinha}}, \ and\ \bibinfo {author}
  {\bibfnamefont {L.}~\bibnamefont {Santos}},\ }\href {\doibase
  10.1103/PhysRevLett.119.050403} {\bibfield  {journal} {\bibinfo  {journal}
  {Phys. Rev. Lett.}\ }\textbf {\bibinfo {volume} {119}},\ \bibinfo {pages}
  {050403} (\bibinfo {year} {2017})}\BibitemShut {NoStop}%
\bibitem [{\citenamefont {Haldane}(1981)}]{haldane1981luttinger}%
  \BibitemOpen
  \bibfield  {author} {\bibinfo {author} {\bibfnamefont {F.}~\bibnamefont
  {Haldane}},\ }\href@noop {} {\bibfield  {journal} {\bibinfo  {journal}
  {Journal of Physics C: Solid State Physics}\ }\textbf {\bibinfo {volume}
  {14}},\ \bibinfo {pages} {2585} (\bibinfo {year} {1981})}\BibitemShut
  {NoStop}%
\bibitem [{\citenamefont {Pitaevskii}\ and\ \citenamefont
  {Stringari}(1991)}]{pitaevskii1991uncertainty}%
  \BibitemOpen
  \bibfield  {author} {\bibinfo {author} {\bibfnamefont {L.}~\bibnamefont
  {Pitaevskii}}\ and\ \bibinfo {author} {\bibfnamefont {S.}~\bibnamefont
  {Stringari}},\ }\href@noop {} {\bibfield  {journal} {\bibinfo  {journal}
  {Journal of low temperature physics}\ }\textbf {\bibinfo {volume} {85}},\
  \bibinfo {pages} {377} (\bibinfo {year} {1991})}\BibitemShut {NoStop}%
\bibitem [{\citenamefont {Aleiner}\ \emph {et~al.}(2010)\citenamefont
  {Aleiner}, \citenamefont {Altshuler},\ and\ \citenamefont
  {Shlyapnikov}}]{aleiner2009finite}%
  \BibitemOpen
  \bibfield  {author} {\bibinfo {author} {\bibfnamefont {I.}~\bibnamefont
  {Aleiner}}, \bibinfo {author} {\bibfnamefont {B.}~\bibnamefont {Altshuler}},
  \ and\ \bibinfo {author} {\bibfnamefont {G.}~\bibnamefont {Shlyapnikov}},\
  }\href {\doibase 10.1038/nphys1758} {\bibfield  {journal} {\bibinfo
  {journal} {Nature Physics}\ }\textbf {\bibinfo {volume} {6}},\ \bibinfo
  {pages} {900} (\bibinfo {year} {2010})}\BibitemShut {NoStop}%
\bibitem [{\citenamefont {Landau}\ \emph {et~al.}(1980)\citenamefont {Landau},
  \citenamefont {Lifshitz},\ and\ \citenamefont
  {Pitaevskii}}]{landau1980statistical}%
  \BibitemOpen
  \bibfield  {author} {\bibinfo {author} {\bibfnamefont {L.~D.}\ \bibnamefont
  {Landau}}, \bibinfo {author} {\bibfnamefont {E.~M.}\ \bibnamefont
  {Lifshitz}}, \ and\ \bibinfo {author} {\bibfnamefont {L.}~\bibnamefont
  {Pitaevskii}},\ }\href@noop {} {\enquote {\bibinfo {title} {Statistical
  physics, part i},}\ } (\bibinfo {year} {1980})\BibitemShut {NoStop}%
\bibitem [{\citenamefont {Giamarchi}(2003)}]{Giamarchi2003}%
  \BibitemOpen
  \bibfield  {author} {\bibinfo {author} {\bibfnamefont {T.}~\bibnamefont
  {Giamarchi}},\ }\href@noop {} {\emph {\bibinfo {title} {Quantum Physics in
  One Dimension}}}\ (\bibinfo  {publisher} {Oxford University Press},\ \bibinfo
  {year} {2003})\BibitemShut {NoStop}%
\bibitem [{\citenamefont {Sutherland}(2004)}]{sutherland2004beautiful}%
  \BibitemOpen
  \bibfield  {author} {\bibinfo {author} {\bibfnamefont {B.}~\bibnamefont
  {Sutherland}},\ }\href@noop {} {\emph {\bibinfo {title} {Beautiful models: 70
  years of exactly solved quantum many-body problems}}}\ (\bibinfo  {publisher}
  {World Scientific Publishing Co Inc},\ \bibinfo {year} {2004})\BibitemShut
  {NoStop}%
\bibitem [{\citenamefont {Cazalilla}\ \emph {et~al.}(2011)\citenamefont
  {Cazalilla}, \citenamefont {Citro}, \citenamefont {Giamarchi}, \citenamefont
  {Orignac},\ and\ \citenamefont {Rigol}}]{cazalilla2011one}%
  \BibitemOpen
  \bibfield  {author} {\bibinfo {author} {\bibfnamefont {M.}~\bibnamefont
  {Cazalilla}}, \bibinfo {author} {\bibfnamefont {R.}~\bibnamefont {Citro}},
  \bibinfo {author} {\bibfnamefont {T.}~\bibnamefont {Giamarchi}}, \bibinfo
  {author} {\bibfnamefont {E.}~\bibnamefont {Orignac}}, \ and\ \bibinfo
  {author} {\bibfnamefont {M.}~\bibnamefont {Rigol}},\ }\href@noop {}
  {\bibfield  {journal} {\bibinfo  {journal} {Reviews of Modern Physics}\
  }\textbf {\bibinfo {volume} {83}},\ \bibinfo {pages} {1405} (\bibinfo {year}
  {2011})}\BibitemShut {NoStop}%
\bibitem [{\citenamefont {Yin}\ \emph {et~al.}(2017)\citenamefont {Yin},
  \citenamefont {Ho},\ and\ \citenamefont {Cui}}]{yin2017majorana}%
  \BibitemOpen
  \bibfield  {author} {\bibinfo {author} {\bibfnamefont {X.}~\bibnamefont
  {Yin}}, \bibinfo {author} {\bibfnamefont {T.-L.}\ \bibnamefont {Ho}}, \ and\
  \bibinfo {author} {\bibfnamefont {X.}~\bibnamefont {Cui}},\ }\href@noop {}
  {\bibfield  {journal} {\bibinfo  {journal} {arXiv preprint arXiv:1711.08765}\
  } (\bibinfo {year} {2017})}\BibitemShut {NoStop}%
\bibitem [{\citenamefont {Lieb}\ and\ \citenamefont
  {Liniger}(1963)}]{lieb1963exactI}%
  \BibitemOpen
  \bibfield  {author} {\bibinfo {author} {\bibfnamefont {E.~H.}\ \bibnamefont
  {Lieb}}\ and\ \bibinfo {author} {\bibfnamefont {W.}~\bibnamefont {Liniger}},\
  }\href {\doibase 10.1103/PhysRev.130.1605} {\bibfield  {journal} {\bibinfo
  {journal} {Phys. Rev.}\ }\textbf {\bibinfo {volume} {130}},\ \bibinfo {pages}
  {1605} (\bibinfo {year} {1963})}\BibitemShut {NoStop}%
\bibitem [{\citenamefont {Gaudin}(1967)}]{gaudin1967systeme}%
  \BibitemOpen
  \bibfield  {author} {\bibinfo {author} {\bibfnamefont {M.}~\bibnamefont
  {Gaudin}},\ }\href@noop {} {\bibfield  {journal} {\bibinfo  {journal}
  {Physics Letters A}\ }\textbf {\bibinfo {volume} {24}},\ \bibinfo {pages}
  {55} (\bibinfo {year} {1967})}\BibitemShut {NoStop}%
\bibitem [{\citenamefont {Yang}(1967)}]{yang1967some}%
  \BibitemOpen
  \bibfield  {author} {\bibinfo {author} {\bibfnamefont {C.}~\bibnamefont
  {Yang}},\ }\href@noop {} {\bibfield  {journal} {\bibinfo  {journal} {Phys.
  Rev. Lett.}\ }\textbf {\bibinfo {volume} {19}},\ \bibinfo {pages} {1312}
  (\bibinfo {year} {1967})}\BibitemShut {NoStop}%
\bibitem [{\citenamefont {Sachdev}(2011)}]{sachdev2011quantum}%
  \BibitemOpen
  \bibfield  {author} {\bibinfo {author} {\bibfnamefont {S.}~\bibnamefont
  {Sachdev}},\ }\href@noop {} {\emph {\bibinfo {title} {Quantum phase
  transitions}}}\ (\bibinfo  {publisher} {Cambridge university press},\
  \bibinfo {year} {2011})\BibitemShut {NoStop}%
\bibitem [{\citenamefont {McGuire}(1964)}]{mcguire1964study}%
  \BibitemOpen
  \bibfield  {author} {\bibinfo {author} {\bibfnamefont {J.~B.}\ \bibnamefont
  {McGuire}},\ }\href@noop {} {\bibfield  {journal} {\bibinfo  {journal}
  {Journal of Mathematical Physics}\ }\textbf {\bibinfo {volume} {5}},\
  \bibinfo {pages} {622} (\bibinfo {year} {1964})}\BibitemShut {NoStop}%
\bibitem [{\citenamefont {Hu}\ \emph {et~al.}(2007)\citenamefont {Hu},
  \citenamefont {Liu},\ and\ \citenamefont {Drummond}}]{hu2007phase}%
  \BibitemOpen
  \bibfield  {author} {\bibinfo {author} {\bibfnamefont {H.}~\bibnamefont
  {Hu}}, \bibinfo {author} {\bibfnamefont {X.-J.}\ \bibnamefont {Liu}}, \ and\
  \bibinfo {author} {\bibfnamefont {P.~D.}\ \bibnamefont {Drummond}},\
  }\href@noop {} {\bibfield  {journal} {\bibinfo  {journal} {Phys. Rev.
  {L}ett.}\ }\textbf {\bibinfo {volume} {98}},\ \bibinfo {pages} {070403}
  (\bibinfo {year} {2007})}\BibitemShut {NoStop}%
\bibitem [{\citenamefont {Guan}\ \emph {et~al.}(2007)\citenamefont {Guan},
  \citenamefont {Batchelor}, \citenamefont {Lee},\ and\ \citenamefont
  {Bortz}}]{guan2007phase}%
  \BibitemOpen
  \bibfield  {author} {\bibinfo {author} {\bibfnamefont {X.-W.}\ \bibnamefont
  {Guan}}, \bibinfo {author} {\bibfnamefont {M.}~\bibnamefont {Batchelor}},
  \bibinfo {author} {\bibfnamefont {C.}~\bibnamefont {Lee}}, \ and\ \bibinfo
  {author} {\bibfnamefont {M.}~\bibnamefont {Bortz}},\ }\href@noop {}
  {\bibfield  {journal} {\bibinfo  {journal} {Physical Review B}\ }\textbf
  {\bibinfo {volume} {76}},\ \bibinfo {pages} {085120} (\bibinfo {year}
  {2007})}\BibitemShut {NoStop}%
\bibitem [{\citenamefont {Girardeau}(1960)}]{Girardeau1960}%
  \BibitemOpen
  \bibfield  {author} {\bibinfo {author} {\bibfnamefont {M.}~\bibnamefont
  {Girardeau}},\ }\href@noop {} {\bibfield  {journal} {\bibinfo  {journal} {J.
  Math. Phys.}\ }\textbf {\bibinfo {volume} {1}},\ \bibinfo {pages} {516}
  (\bibinfo {year} {1960})}\BibitemShut {NoStop}%
\end{thebibliography}%

\end{document}